\documentclass[aps,twocolumn]{revtex4}
\usepackage{graphics}
\usepackage{epsfig}

\begin{document}

\title{Noise-induced cooperative dynamics and its control in coupled 
neuron models}
\author {B. Hauschildt$^1$} 
\email{hauschildt@nlds.physik.tu-berlin.de}
\author{N.B. Janson$^2$}
\email{n.b.janson@lboro.ac.uk}
\author{A. Balanov$^{1,3}$}
\email{alexander.balanov@nottingham.ac.uk}
\author{E. Sch\"oll$^1$}
\email{schoell@physik.tu-berlin.de}
\affiliation {$^1$Institut f{\"u}r Theoretische
Physik, Technische Universit{\"a}t Berlin, Hardenbergstrasse 36, 
D-10623 Berlin, Germany\\
$^2$Department of Mathematical Sciences, Loughborough University,
Loughborough, Leicestershire, LE11 3TU, UK\\
$^3$School of Physics and Astronomy,
University of Nottingham,
University Park, 
Nottingham, NG7 2RD, UK}

\date{\today}

\begin{abstract} 
We investigate feedback control of the cooperative dynamics of two coupled 
neural oscillators that is induced merely by external noise. The 
interacting neurons are modelled as FitzHugh-Nagumo systems with 
parameter values at which no autonomous oscillations occur, and each 
unit is forced by its own source of random fluctuations. Application 
of delayed feedback to only one of two subsystems is shown to be able
to change coherence and timescales of noise-induced oscillations either
in the given subsystem, or globally. It is also able to induce or to
suppress stochastic synchronization under certain
conditions.\end{abstract}

\maketitle

\section{Introduction}

Neural systems in many cases are characterized by oscillatory behavior 
\cite{BRA94, LU95, EGU00}, which is often quite complicated 
\cite{SOF93}. It has been shown that neural oscillatory dynamics
can have different origins, being either self-sustained \cite{HOD52}, 
or induced by random fluctuations alone \cite{JUN98,BAD05}. 
These oscillations can also be multimodal, i.e., consisting of several 
components with different prominent timescales. For example the 
thalamocortical relay neurons can generate either spindle or $\delta$ 
oscillations \cite{WAN94}, whereas the electro-receptors in a paddle 
fish are able to generate biperiodic oscillations \cite{NEI01}. Coupled 
neurons are able to demonstrate synchronization, which plays a very 
important role in neurodynamics, having either 
constructive, or destructive effects depending on the circumstances. 

On one hand, the ensembles of different
neurons can be synchronized in order to process biological information, 
i.e., this 
synchronization might be beneficial for a more efficient data transmission 
\cite{SAM04,BEN04}. On the other hand, these synchronized neurons can 
induce a  regular, rhythmic activity, which is believed to play a 
crucial role in the emergence of pathological rhythmic brain activity in 
Parkinson's disease, essential tremor, and 
epilepsy \cite{TAS98,GRO02}. In both situations, synchronization 
phenomena occur spontaneously, and the mechanisms behind them are the 
subjects of  intensive research \cite{GRO02, BEN04}. Hence, the 
development of techniques that would allow one to manipulate the neural 
synchrony is an important clinical problem. 

Starting with the work of 
Ott, Grebogi and Yorke \cite{OTT90}, a variety of methods for the 
control of irregular behavior have been developed in the last 15 years 
\cite{SCH99c,BOC00}. Recently, a number of methods have been proposed 
for suppression of synchrony in the arrays of coupled oscillators  in 
which oscillations are self-sustained \cite{ROS04,POP05}. However, the 
problem of control of the dynamics in coupled systems where 
oscillations are induced merely by random fluctuations is still open. 

In this paper we investigate the possibility of using for this purpose 
the delayed feedback scheme introduced by Pyragas \cite{PYR92}: 
it constructs a control force from the difference 
between the current state of the system and its state some $\tau$ time 
units ago. This method is known as \textit{time-delay 
autosynchronization} (TDAS). It has been applied to control 
of deterministic chaos in a wide range of systems including spatially 
extended models, e.g. \cite{BAB02,BEC02,UNK03}. 
It was also demonstrated that it can be used to
control the coherence and the timescales of noise-induced oscillations 
in a single system \cite{JAN04,BAL04,POM05a}. This theoretical prediction
has recently been verified experimentally in an electrochemical
oscillator system \cite{SAN06}.
The main aim of the present work is to extend time-delayed feedback control
of noise-induced dynamics to {\em coupled} excitable systems, and investigate if
{\em local} control applied to a subsystem can allow one to steer the 
{\it  global} cooperative dynamics in a system of coupled neural oscillators. 
In particular, we are interested in the study of the 
effects of delayed feedback on the {\em synchrony} properties in coupled 
neuron systems.  

The paper is organized as follows. In Section~\ref{sec:system} 
we introduce the model system used, and discuss the properties of 
their cooperative behavior without control. In 
Section~\ref{sec:control} we study the effects of local delayed feedback 
control  on the global behavior of  coupled systems. In 
Section~\ref{sec:sum}  the results are summarized and the conclusions 
are drawn. 

\section{\label{sec:system} Global dynamics of two coupled neural oscillators}

In order to grasp the complicated interaction between billions of neurons in large
neural networks, those are often lumped into groups of neural populations 
each of which can be represented as an effective excitable element that is mutually 
coupled to the other elements 
\cite{ROS04,POP05}. In this sense the simplest model which may reveal features 
of interacting neurons consists of two coupled neural oscillators. Each of these will be 
represented by a simplified FitzHugh-Nagumo system, which is often used as a 
paradigmatic generic model for neurons, or more generally, excitable systems
\cite{LIN04}. Here we use two FitzHugh-Nagumo systems with substantially different
intrinsic timescales, and parameters corresponding to the excitable regime. 
%In order to gain insight into the complex nonlinear dynamics of interacting neurons
%we adopt a simple generic model. 
Before attempting to control their global dynamics 
with locally applied feedback, we will first study the dynamics of the uncontrolled
coupled system.
The dynamical equations are given by:
\begin{eqnarray}	  
\label{eqn:fhnc1}     \epsilon_{1}\, \dot{x}_{1} & = & 
x_{1}-\frac{x_{1}^{3}}{3}-y_{1}+C\,\left( 
x_{2}-x_{1}\right),\nonumber\\     \dot{y}_{1} & = & 
x_{1}+a+D_{1}\,\xi_1 (t),\\     \label{eqn:fhnc2}      \epsilon_{2}\, 
\dot{x}_{2} & = & x_{2}-\frac{x_{2}^{3}}{3}-y_{2}+C\,\left( 
x_{1}-x_{2}\right),\nonumber\\      \dot{y}_{2} & = & 
x_{2}+a+D_{2}\,\xi_2 (t),\end{eqnarray}
where subsystems Eqs. 
(\ref{eqn:fhnc1}) and (\ref{eqn:fhnc2}) represent two different neurons, 
$x_i$ ($i=1,2$) describing the transmembrane voltages and $y_i$ modelling 
the behavior of several physical quantities related to electrical 
conductances of the relevant ion currents across the respective 
membranes.  Here $a$ is a bifurcation parameter whose value defines 
whether the system is excitable or demonstrates periodic firing 
(autonomous oscillations), 
$\epsilon_{1}$ and $\epsilon_{2}$ are positive parameters that 
are usually chosen to be much 
smaller than unity, $\xi_{1}$ and $\xi_{2}$ are independent sources of 
Gaussian white
noise with zero mean and unity variance, $D_{1}$ and $D_{2}$ are noise 
intensities.

The synaptic coupling between two neurons 
is modelled as a diffusive coupling considered for simplicity to be
symmetric \cite{LIL94,PIN00,DEM01}. The coupling 
strength $C$ summarizes how information is distributed between neurons.

We shall restrict our analysis to the range of the
parameter values where without noise each of the two subsystems
exhibits excitability 
with only one attractor in the form of a stable fixed point. 
 Noise sources $\xi_1$ and $\xi_2$ model random
inputs that represent integral signals coming from the part of the
neural network or of the environment with which the neuron is
connected. Since the neurons can be coupled to different parts of the 
neural network or of the environment, the noise intensities in the two
systems can be quite different. 

\begin{figure}
\epsfig{file=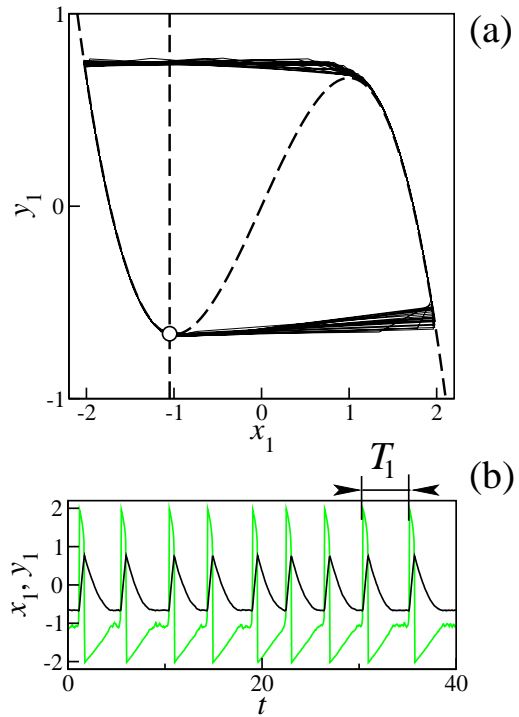,angle=0,width=7cm}
\caption{\label{fig:pp}(color online) 
(a) Phase portrait and (b)
realizations of $x_1$ (grey, green online) and $y_1$ (black) time series
in a single FitzHugh-Nagumo system Eq. (\ref{eqn:fhnc1}). In (a) dashed
lines are null-clines and the fixed point is shown by a white circle.
Parameters: $a$$=$$1.05$, $\epsilon_{1}$$=$$0.005$, $D_1=0.02$. 
}\end{figure}

\subsection{\label{sec:oneneuron} Features of a single neuron model}

Let us illustrate the dynamics of a single neuron model  
by considering an uncoupled subsystem Eqs. (\ref{eqn:fhnc1})
($C$$=$$0$) under the influence of noise. We arbitrarily fix  
$D_1$$=$$0.02$, and also set $a=1.05$, $\epsilon_{1}$$=$$0.005$. 
In Fig. \ref{fig:pp}(a) dashed lines show the null-clines of 
Eq. (\ref{eqn:fhnc1}) that intersect at a fixed point marked by a white 
circle. The phase point that is initially placed at the fixed point 
stays in its close vicinity if the applied random perturbation remains 
small.  However, if the perturbation is larger than some threshold 
value, the phase point makes a large excursion in the phase space 
before returning to the vicinity of the fixed point again. In Fig. 
\ref{fig:pp}(a) the black solid line illustrates a phase portrait and in Fig. 
\ref{fig:pp}(b) realizations of $x_1$ and $y_1$ time series from Eqs. 
(\ref{eqn:fhnc1}) are shown. The motion of the phase point 
consists of two stages: an activation time during which the system waits 
for a sufficiently large perturbation before it can make an excursion, 
and the excursion itself. The excursion time is almost completely defined 
by the deterministic properties of the system and is hardly 
influenced by noise. 

On the contrary, the activation time is completely 
determined by the properties of noise if all other parameters are fixed: 
the stronger the noise, the smaller the activation time and the larger
the mean frequency of noise-induced oscillations is. Thus, the noise 
strengths $D_1$ and $D_2$ control the average frequencies of
noise-induced oscillations in the systems Eqs. (\ref{eqn:fhnc1}) and 
(\ref{eqn:fhnc2}), respectively, and the difference between them 
defines the mean frequency 
detuning between the systems. 

\subsection{\label{sec:means}
Means for characterizing cooperative dynamics}

Before studying the effects of delayed feedback control on the model 
Eqs. (\ref{eqn:fhnc1}) and (\ref{eqn:fhnc2}), let us examine the basic 
features of cooperative dynamics of two systems in which oscillations 
are induced merely by noise. 

The cooperative dynamics of a system of coupled stochastic 
oscillators can be
characterized differently depending on the feature of interest. 
The most popular features are timescales involved, degree 
of order in each partial subsystem and in the system of coupled
oscillators as a whole, and
the degree of synchronism between the subsystems. 
To quantitatively characterize each  feature of interest, a number 
of criteria can be used, and here we will choose those that
seem to suit best our purposes. 

{\bf Timescales.}
The Fourier power spectral density, to which in the following we
will refer as {\em spectrum} for brevity, seems to be the most universal
and sensitive tool that allows one to fully reveal the frequency
content of random oscillations and thus characterize the timescales
involved. The central frequencies of the highest spectral peaks 
will characterize the timescales involved. 

Another convenient and less computationally expensive way to characterize 
the timescales of oscillations is to introduce the interspike intervals
(ISI) $T_1$ and $T_2$ for the two systems from their 
realizations $x_1(t)$ and $x_2(t)$, respectively, as shown in Fig. 1(b). 
The average ISIs $\langle T_i \rangle$, $i=1,2$,
will also characterize the timescales of two systems. 

{\bf Coherence.}
Generally, the width of the spectral peak can serve as an indication of 
the coherence of the oscillations:
the narrower the peak is, the more coherent the oscillations are. 
However, as we will see below, the spectra of the observed oscillations
have several distinguishable peaks with comparable heights 
placed at incommensurate frequencies
(i.e., at frequencies that are not multiples of each other), and all 
peaks have different widths. It is not obvious the width of which peak
should be taken as a measure of coherence, and thus the peak width
does not represent an unambiguous criterion here and will not be used.

Another measure of coherence of oscillations is the correlation time $t_{cor}$.
It is also not unambiguous because it can be introduced in several ways. 
Here we will use the following method which seems the most universal:
an autocorrelation function $\Psi(s)=\langle[x(t-s)-\langle 
x\rangle][x(t)-\langle x\rangle]\rangle$ will be calculated from the 
simulated realizations $x(t)$, and $t_{cor}$ will be introduced as
\begin{equation}
t_{cor}=\frac{1}{\sigma^{2}}\int_{0}^{\infty}\vert\Psi(s)\vert 
\,ds,\end{equation}
where $\sigma^{2}=$ $\Psi(0)$ is the variance of $x(t)$. 
The larger $t_{cor}$ is, the more regular $x(t)$ is. 

Timescales and coherence can be introduced for each subsystem
separately and then compared, or for some variable characterizing
the state of the system as a whole. Thus, we will
estimate the statistical characteristics of the variables
$x_1$ and $x_2$, and of the global state variable $x_\Sigma=x_1+x_2$.

{\bf Synchronization.}
Finally, we need to characterize the synchronization 
between the two coupled oscillators. Most generally, synchronization
means an adjustment of timescales of oscillations in systems
due to the interaction between them: if the timescales in the 
uncoupled systems are not rationally related, introduction of
coupling can shift the timescales to make their ratio closer
to a rational number $n$$:$$m$, where $n$ and $m$ are integers. 
This phenomenon is usually referred to as $n$$:$$m$ 
frequency synchronization, and its suitable measure 
would be the closeness of the ratio of average 
ISIs $\langle T_1 \rangle$$/$$\langle T_2 \rangle$ to the chosen 
rational number $n$$:$$m$. Note that frequency
synchronization is associated with 
the time-averaged behavior of the coupled oscillators.

A closely related, but not identical, phenomenon, which is usually 
called phase synchronization,
is associated with instantaneous coordination between the interacting
systems. It requires the definition of phases $\varphi_1(t)$ and 
$\varphi_2(t)$
for each oscillator and comparison between them. 
In our system the spiky nature of oscillations allows one to introduce the
phase for each system as:
\begin{equation}
\varphi_{j}(t)=2\pi\frac{t-t_{i-1}}{t_{i}-t_{i-1}}+2\pi(i-1),\,\,j=1,2,
...,\end{equation}
where $t_{i}$ is the time at which we observe a spike
in the respective system's realization. 

We define
$n$$:$$m$ phase synchronization to occur if the
phase difference 
\begin{equation}
\Delta\varphi_{n,m}(t)=\varphi_{1}(t)-\frac{m}{n}\varphi_{2}(t),
\end{equation}
exhibits horizontal plateaus of sufficient duration. 
Usually, if $n$$:$$m$ synchronization takes place, $\Delta\varphi_{n,m}(t)$ 
demonstrates plateaus occasionally interrupted by $2\pi$ jumps. 
On the plateaus, $\Delta\varphi_{n,m}(t)$ usually oscillates around
some local average level. As time grows, $\Delta\varphi_{n,m}(t)$
drifts to plus or minus infinity. 

In \cite{Ros_Moss_book} several measures to characterize
phase synchronization were introduced. Here, we choose to estimate
the synchronization index 
\begin{equation}
\gamma_{n,m}=\sqrt{\langle \cos \Delta\varphi_{n,m}(t) \rangle^2 +
\langle \sin \Delta\varphi_{n,m}(t) \rangle^2}. 
\end{equation}
$\gamma_{n,m}$ can vary between $0$ (no synchronization) and $1$ (perfect
$n$$:$$m$ phase synchronization). 
Note that even if the ratio of average ISIs is close or even equal
to some rational number $n$$:$$m$, i.e., frequency synchronization
takes place, phase synchronization does not necessarily occur,
and the synchronization index might be close to zero. 

\subsection{\label{sec:features}
Cooperative noise-induced dynamics in two coupled neurons: 
timescales and coherence}

All results in this paper are presented for $a$$=$$1.05$, 
$\epsilon_{1}$$=$$0.005$,
$\epsilon_{2}$$=$$0.1$, and $D_2=0.09$. Mean frequency detuning will
be determined by the choice of $D_1$. Note that 
$\epsilon_1 \ne \epsilon_2$, i.e. the systems are not identical, 
therefore at $D_1=D_2$ the mean frequencies
of oscillations in two uncoupled ($C=0$) systems will be different. 

We need to find out how the cooperative dynamics of the 
interacting systems
changes depending on coupling strength $C$ and on the mean frequency 
detuning defined by $D_1$. 

\begin{figure}
\epsfig{file=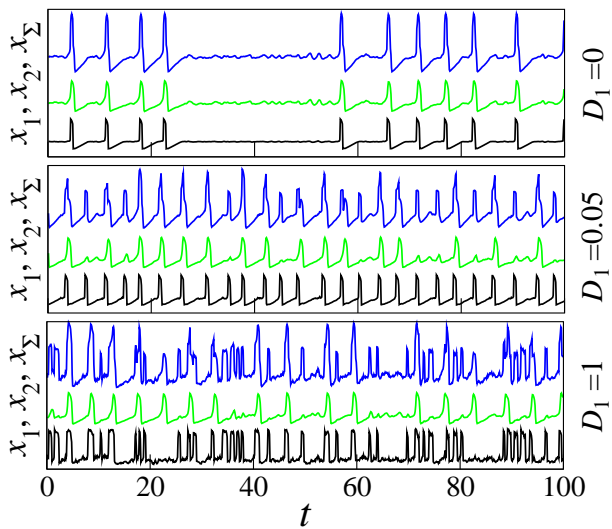,angle=0,width=8cm}
\caption{\label{fig:fig_C_0.07}(color online) 
Realizations of $x_{1}$ (lower), $x_{2}$ (middle) and $x_{\Sigma}=x_1+x_2$ 
(upper trace) of noise-induced oscillations in two coupled
FitzHugh-Nagumo systems Eqs. (\ref{eqn:fhnc1}), (\ref{eqn:fhnc2})
for various noise intensities $D_{1}$ in the first system for
$C$$=$$0.07$. }\end{figure}

We first fix the coupling strength $C$ at $0.07$, and change $D_1$. 
Fig. \ref{fig:fig_C_0.07} shows the realizations 
$x_{1}$, $x_{2}$ and $x_{\Sigma}$
of noise-induced oscillations. At $D_{1}$$=$$0$ the first
subsystem, whose variables are denoted by subscript 1 in Eq. (\ref{eqn:fhnc1}), has
no independent dynamics. But due to the coupling with the second
subsystem, it demonstrates forced oscillations whose properties are
completely defined by those of the second subsystem. The respective
realizations of $x_{1}$ and $x_{2}$ demonstrate excellent synchrony,
so that each spike in $x_{2}$ causes a spike in $x_{1}$ that occurs 
simultaneously (Fig. \ref{fig:fig_C_0.07} ($D_1$$=$$0$)). 

\begin{figure}
\epsfig{file=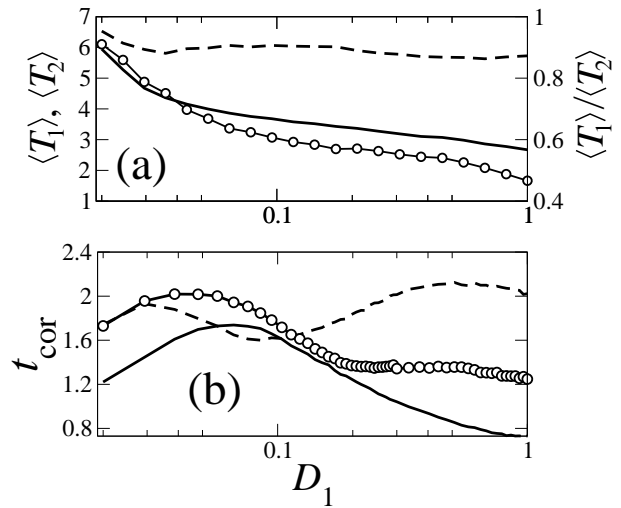,angle=0,width=8cm}
\caption{\label{fig:fhnc_spec_tcor_D1}
Timescales and coherence of noise-induced oscillations in two coupled
FitzHugh-Nagumo systems Eqs. (\ref{eqn:fhnc1}) and (\ref{eqn:fhnc2})
{\em vs} $D_{1}$ at $C$$=$$0.07$. 
(a) Average ISIs $\langle T_1 \rangle$ (solid line), 
$\langle T_2 \rangle$ (dashed line) and
their ratio (circles). (b)
Correlation time $t_{cor}$ obtained from $x_{1}$ (solid line),
$x_{2}$ (dashed line) and $x_{\Sigma}$ (circles).
See text for details.}\end{figure}

At $D_1 \ne 0$ the first subsystem acquires its own dynamics with the respective
independent timescale. Now each time one of the subsystems produces a 
noise-induced spike, due to coupling the other subsystem is prompted 
to spike, too: it does not necessarily emit a spike,
but the spiking probability grows slightly.
As a result, both subsystems are likely to spike slightly more frequently 
(Fig. \ref{fig:fig_C_0.07} 
($D_1$$=$$0.05$)) This is reflected by the decrease of 
respective average ISIs in Fig. 
\ref{fig:fhnc_spec_tcor_D1}(a). 
As $D_1$ grows further, the mean frequency of
spiking in the first subsystem grows in agreement with \cite{Lind_CR_99}.
However, coupling is small here, so the second subsystem only rarely
responds with a spike to the spike in the first subsystem. 
As a result, while the spiking frequency in the first
subsystem is further increased, the second subsystem's frequency
stays almost constant (Fig. \ref{fig:fig_C_0.07} 
($D_1$$=$$1$) and \ref{fig:fhnc_spec_tcor_D1}(a)).

The continuous change of 
timescales and of coherence of the noise-induced dynamics
with $D_1$ is illustrated in Fig.~\ref{fig:fhnc_spec_tcor_D1}.
Here, average ISIs (a) and $t_{cor}$ (b) are shown. The latter are estimated
from $x_1$ and $x_2$ thus quantifying the local dynamics, and
from $x_{\Sigma}$$=$$x_1$$+$$x_2$ thus characterizing the global
behavior. 

All
three graphs for $t_{cor}$ show clear maxima which can be regarded as 
occurrence of coherence resonance (CR) \cite{PIK97}. However, in
different systems CR occurs at different noise intensities $D_1$, 
and the mutual coupling between the two systems leads to the 
occurrence of two maxima in $t_{cor}$ calculated from $x_2$. 
%which leads to the occurrence of two maxima in $t_{cor}$ calculated 
%from $x_{\Sigma}$.

\begin{figure}
\epsfig{file=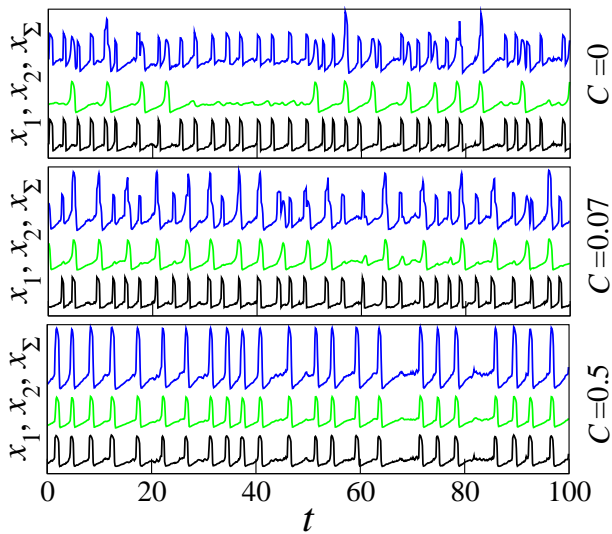,angle=0,width=8cm}
\caption{\label{fig:fig_D1_0.25}(color online) 
Realizations of $x_{1}$ (lower), $x_{2}$ (middle) and $x_{\Sigma}$ (upper trace)
of noise-induced oscillations in two coupled
FitzHugh-Nagumo systems Eqs. (\ref{eqn:fhnc1}) and (\ref{eqn:fhnc2})
for various coupling strengths $C$ at
$D_1$$=$$0.25$. }\end{figure}

Now consider how noise-induced dynamics changes 
with variation of the coupling $C$ between neurons. 
We choose  $D_{1}$$=$$0.25$, so that without
coupling ($C$$=$$0$) the oscillations in the two systems have essentially 
different timescales with $\langle T_1 \rangle \approx 3.25$
and $\langle T_2 \rangle \approx 8.1$. 
%(Fig. \ref{fig:fhnc_spec_tcor_D1}(a)). 
At $C$$=$$0$ the two subsystems oscillate independently 
(Fig. \ref{fig:fig_D1_0.25}, top panel).
As the coupling is increased from zero, the subsystems 
start to experience each others' influence: each time one
system spikes, the other is prompted to spike, too
(Fig. \ref{fig:fig_D1_0.25}, middle panel). This results
in timescales moving closer (Fig. \ref{fig:fhnc_spec_tcor_D1}(a)). 
As the coupling grows further, the two subsystems spike more
simultaneously (Fig. \ref{fig:fig_D1_0.25}, bottom panel),
and their average ISIs tend to coincide.   
The latter can serve as an evidence for stochastic $1$$:$$1$ frequency 
synchronization 
\cite{HAN99}, which will be discussed in more detail below. 

\begin{figure}
\epsfig{file=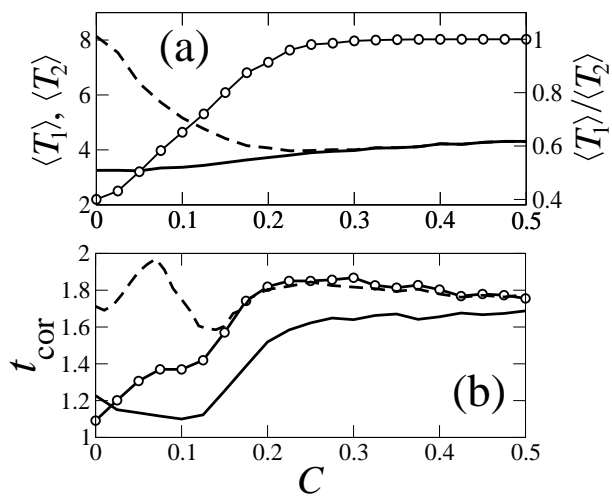,angle=0,width=8cm}
\caption{\label{fig:fhnc_spec_tcor_C}
Timescales and coherence of noise-induced oscillations in two coupled
FitzHugh-Nagumo systems Eqs. (\ref{eqn:fhnc1}) and (\ref{eqn:fhnc2})
{\em vs} coupling strength $C$ at
$D_1$$=$$0.25$. 
(a) Average ISIs $\langle T_1 \rangle$ (solid line), 
$\langle T_2 \rangle$ (dashed line) and their ratio
(circles).
 (b)
Correlation time $t_{cor}$ obtained from $x_{1}$ (solid line),
$x_{2}$ (dashed line) and $x_{\Sigma}$ (circles).
See text for details.
}\end{figure}
  
The full dependence of ISIs and $t_{cor}$ on $C$ is shown in 
Fig.~\ref{fig:fhnc_spec_tcor_C}.  
In the absence of coupling, the two systems randomly oscillate,
being independent of each other, hence the coherence of the
sum signal is less than the coherences of the individual signals. 
At large $C>0.2$, the global  coherence becomes equal to the 
coherence of the second system,
which is more ordered individually than its neighbour. 

\subsection{\label{sec:syn}
Synchronization: frequency (phase)
locking and suppression of noise-induced oscillations}

Synchronization phenomena in coupled oscillators with 
noise-induced dynamics were previously considered, e.g. 
in \cite{HAN99,POS02}. 
In contrast to these works, our model consists of essentially 
non-identical subsystems whose dynamics is defined by independent 
sources of noise with different strength, which describes a more 
general class of natural systems.  

\begin{figure}
\epsfxsize=\linewidth
%\epsfbox{FIG/fig_D1_C_all.eps}
\epsfbox{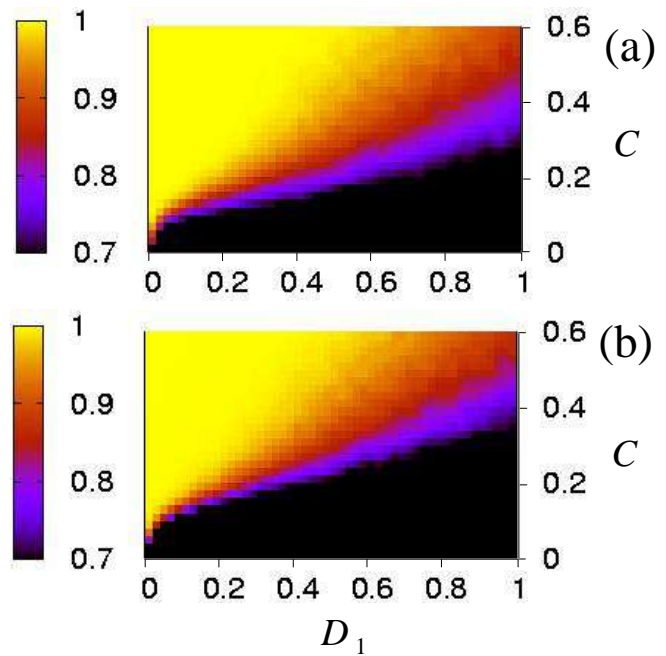}
\caption{\label{fig:fhnc_isi_D1_C} (color online) (a) Ratio of
average interspike intervals 
$\langle T_1 \rangle / \langle T_2 \rangle$ from 
the two systems, and (b) synchronization index $\gamma_{1,1}$ 
{\it vs} the coupling
strength $C$ and noise intensity $D_{1}$. The light (yellow) areas roughly
outline the $1$$:$$1$ (a) frequency and (b) phase synchronization tongues.}
\end{figure}

In this paper we will discuss only $1$$:$$1$ synchronization. 
Since synchronization means an adjustment of timescales 
in interacting systems, the ratio of their average ISIs would serve
as a good tool for its detection. In 
Fig.~\ref{fig:fhnc_isi_D1_C}(a) the ratio 
$\langle T_1 \rangle / \langle T_2 \rangle$ is shown for a range of 
coupling strengths $C$ and of noise intensities $D_{1}$. 
One can clearly see the $1$:$1$
synchronization region (light area), which has a quite recognizable 
tongue-like shape, i.e., the larger the coupling strength, the wider the 
synchronization region with respect to $D_1$ is.

\begin{figure}
\epsfig{file=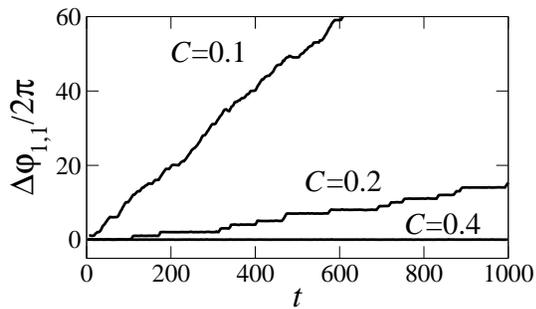,angle=0,width=7cm}
\caption{\label{fig:fhnc_phase11} 
Phase difference $\Delta\varphi_{1,1}$ for the suspected $1$:$1$ 
synchronization at three different values of coupling
strength $C$ and $D_{1}$$=$$0.25$.}\end{figure}

Next, we explore if phase synchronization accompanies the
frequency synchronization. 
In Fig.~\ref{fig:fhnc_phase11} the phase
difference is shown for 
$D_{1}$$=$$0.25$: as $C$ is  
increased from $0.1$ to $0.4$, the plateaus of $\Delta\varphi_{n,m}(t)$
become longer. At $C=0.4$ the coupling is so strong
that the plateau exists (almost) infinitely, which means that
the two systems are well $1$:$1$ phase synchronized. 

The synchronization index $\gamma_{1,1}$ is computed in the
whole range of $D_1$ and $C$ and is shown in Fig. 
\ref{fig:fhnc_isi_D1_C}(b). Inside the whole region of $1$:$1$ 
frequency synchronization where the ISI ratio is close to unity,
the phase synchronization index is close to unity, too.
Thus, phase synchronization occurs together with frequency
synchronization.

\begin{figure} 
\epsfxsize=\linewidth
%\epsfbox{FIG/fig_lock_supp_spec.eps}
\epsfbox{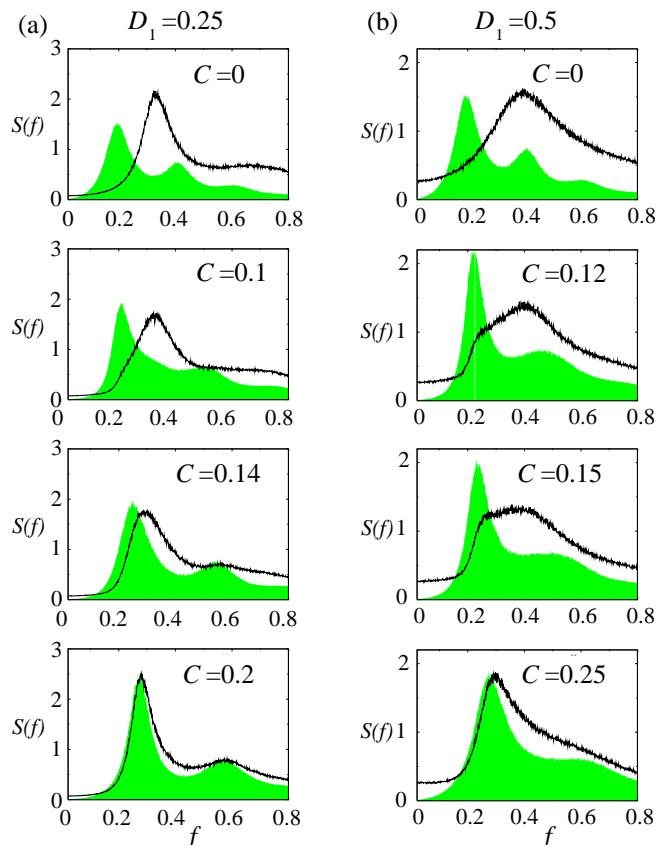}
\caption{\label{fig:fhnc_syn_mech} 
(Color online) The spectra of $x_1$ (black) 
and $x_2$ (shaded grey area, green online) 
are compared as the coupling strength $C$ is 
increased for two different strengths of noise in the first subsystem (a) 
$D_{1}$$=$$0.25$ and (b) $D_{1}$$=$$0.5$. 
In (a) the highest spectral peaks move 
towards each other and coincide: frequency 
(phase) locking. In (b) the highest peak in the spectrum of $x_1$ is 
suppressed, while another peak simultaneously appears and grows at the 
peak frequency of $x_2$: suppresion of natural dynamics.}
\end{figure}

It has been known for a long time that
synchronization can be achieved via at least two different mechanisms,
namely {\it frequency (phase) locking}, and {\it suppression of natural
dynamics}, respectively (see \cite{LAN80} for periodic oscillations and 
\cite{MOS02}
for chaotic and noise-induced oscillations). We found that in our 
model both these
synchronization mechanisms can be realised, depending on how well the
timescales of interacting oscillators were separated from each other when
uncoupled. An example of synchronization via frequency (phase) locking
is illustrated by Fig.~\ref{fig:fhnc_syn_mech}(a), where for
$D_1$$=$$0.25$ the spectra of signal $x_1$ and of $x_2$ are illustrated
for increasing $C$. As $C$ grows, two distinguishable peaks
corresponding to the timescales of the two subsystems move closer and
then merge. Fig.~\ref{fig:fhnc_syn_mech}(b) shows how synchronization is
realised via the suppression of natural dynamics at $D_1$$=$$0.5$. One
can see that the increase of $C$ suppresses one of the spectral peaks,
i.e. one of the timescales of the system (\ref{eqn:fhnc1}). Thus, 
mutually coupled systems with noise-induced spiking are able not only 
to demonstrate mutual synchronization itself, but also to reproduce two 
different synchronization mechanisms, in full analogy with coupled 
self-oscillating systems. 

\section{\label{sec:control}Local delayed feedback control of 
noise-induced cooperative dynamics}

In this section we investigate whether the 
feedback applied only to one of the interacting subsystems, i.e., locally,
is able to manipulate the global properties of the system
of coupled oscillators. This might simulate a realistic 
situation where only a small area of the neural network is 
available for external stimulation. 
In particular, we will investigate if
global timescales, coherence and the strength of synchronization
can be influenced. 

The time-delayed feedback 
control proposed by Pyragas \cite{PYR92} for control of 
deterministic chaos was previously applied for the 
control of noise-induced oscillations in a 
single system with noise-induced dynamics
\cite{JAN04,BAL04,POM05a}. It has been demonstrated that 
it can successfully change the timescales
and coherence of oscillations and is thus a promising 
tool for control of noise-induced phenomena in general. 
 
For our purpose, we apply the time-delayed feedback  
to the first subsystem alone, while the second system
remains freely coupled to it. 
The feedback force $F(t)$ is constructed as follows:
the slow state variable $y_1$ is saved
at the current time $t$ and at a time $(t-\tau)$, their
difference is calculated and multipled by the feedback strength $K$. 
$F(t)$ is then fed back to the $y$-component of the vector field
\begin{eqnarray}	     
\label{eqn:fhnc1_fdbk}
\epsilon_{1}\, \dot{x}_{1} & = &
x_{1}-\frac{x_{1}^{3}}{3}-y_{1}+C\,\left( 
x_{2}-x_{1}\right),\\      \label{eqn:fhncwcontrol}     
\dot{y}_{1} & = & x_{1}+a+K\,[y_{1}(t-\tau)-y_{1}(t)]+D_{1}\,\xi 
(t),\nonumber\end{eqnarray}
where $\tau$ is the time delay and the other parameters are as in
Eqs. (\ref{eqn:fhnc1}). 

We will be guided by the full picture of cooperative dynamics of the 
two mutually coupled subsystems that was revealed in 
Sec.~\ref{sec:system}. We will choose states with different global dynamics 
by choosing pairs of parameters $D_1$ and $C$, and study the effect
of the delayed feedback on each state. 

We select pairs of points ($D_1$, $C$)
(see Fig.~\ref{fig:fhnc_isi_D1_C}) at which 
the two systems are (i) far away from ($D_1$$=$$0.6$, $C$$=$$0.1$), 
(ii) closer to ($D_1$$=$$0.6$, $C$$=$$0.2$), and (iii) almost inside
($D_1$$=$$0.15$, $C$$=$$0.2$)
the $1$$:$$1$ synchronization
region. In Sec. \ref{subsec:moderate_syn} we 
study in detail the case of a moderately synchronized system 
at $D_1$$=$$0.6$ and $C$$=$$0.2$, 
subject to delayed feedback. We reveal the common features
of the feedback effect depending on its parameters $\tau$ and $K$. 

Further on, in Sec. \ref{subsec:all_syn}, we 
study two more cases of systems further from, and closer to,
the $1$$:$$1$ synchronization region
under the delayed feedback action. We compare the effect of the
feedback with its effect on a moderately synchronized system. 

\subsection{\label{subsec:moderate_syn}Control of a moderately
synchronized system}

\begin{figure} 
\epsfxsize=\linewidth
%\epsfbox{FIG/fig_K_tau_D1_0.6_C_0.2.eps}
\epsfbox{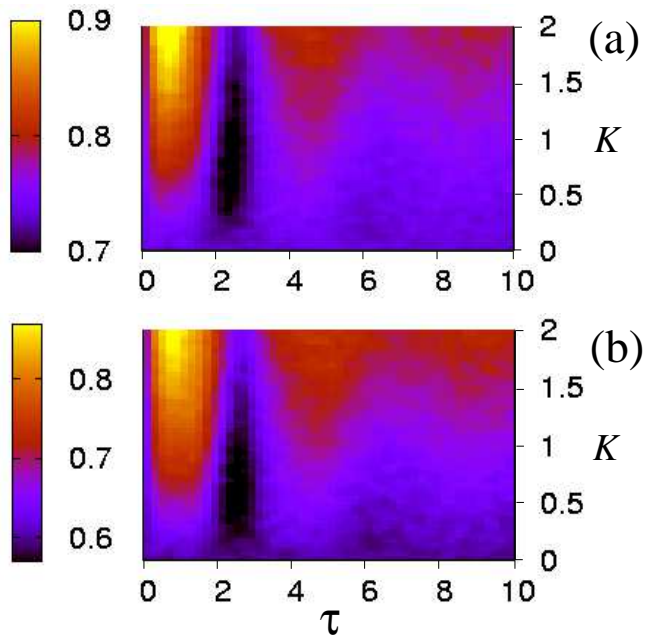}
\caption{\label{fig:fhnc_K_tau_D1_0.6_C_0.2} 
(color online) 
Effect of delayed feedback on frequency and phase 
synchronization betwen the two subsystems at
$D_1$$=$$0.6$ and $C$$=$$0.2$, which corresponds to a
moderate distance from the 
$1$$:$$1$ synchronization tongue  
shown in Fig.~\ref{fig:fhnc_isi_D1_C}. 
(a) Ratio of
average interspike intervals 
$\langle T_1 \rangle / \langle T_2 \rangle$ from 
the two systems and 
(b) synchronization index 
$\gamma_{1,1}$ 
{\it vs} the control strength $K$ and 
the time-delay $\tau$. 
}\end{figure}

Here, we consider subsystems Eqs.~(\ref{eqn:fhnc1_fdbk}) 
and~(\ref{eqn:fhnc2}) 
with $D_1$$=$$0.6$ and $C$$=$$0.2$, under the influence
of the controlling feedback. 
We aim to find out if the feedback can make the subsystems
more, or less, synchronous, and their global dynamics more or less
coherent. In particular, we are interested if perfect 
$1$$:$$1$ synchronization can be induced by the local feedback,
or if the existing synchronization can be destroyed. 
The ratio of ISIs and the synchronization index $\gamma_{1,1}$
are shown by color code 
in Fig. 
\ref{fig:fhnc_K_tau_D1_0.6_C_0.2} 
for a large range of
the values of the feedback delay $\tau$ and strength $K$. 
The lighter areas are associated with the stronger 
$1$$:$$1$ synchronization, and the values at $K$$=$$0$ and 
at $\tau$$=$$0$
characterize the original state of the system without feedback. 
As seen from Fig.~\ref{fig:fhnc_K_tau_D1_0.6_C_0.2}, 
the locally applied delayed feedback is able to
move the system's state closer to 
the $1$$:$$1$ synchronization with suitable 
feedback parameters. 
On the other hand, for $\tau \approx 2.5$ (black area)
$1$$:$$1$ synchronization is suppressed.

\begin{figure}
\epsfig{file=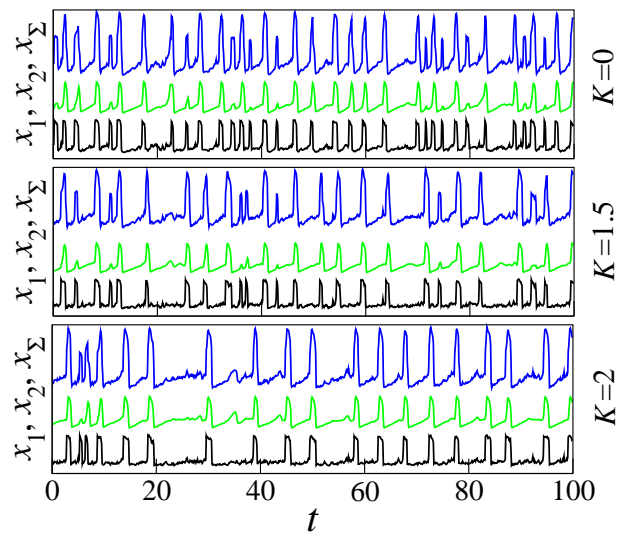,angle=0,width=8cm}
\caption{\label{fig:fig_tau_1}(color online) 
Realizations of $x_{1}$ (lower), $x_{2}$ (middle) and $x_{\Sigma}$ (upper trace)
of noise-induced oscillations in two coupled
FitzHugh-Nagumo systems Eqs.~(\ref{eqn:fhnc1_fdbk}) and (\ref{eqn:fhnc2})
at $D_1$$=$$0.6$ and $C$$=$$0.2$, 
subject to delayed feedback with $\tau=1$
for different values of the feedback strength $K$.
See Fig.~\ref{fig:ISI_K_D1_0.6_C_0.2_tau_1} for reference.}\end{figure}

An illustration of  how realizations $x_1$, $x_2$ and $x_{\Sigma}$
change depending on the feedback strength $K$ as $\tau$$=$$1$ is fixed,
is given in Fig.~\ref{fig:fig_tau_1},
Note that the cut at $\tau$$=$$1$ (Fig. 9) goes through the
region where the strongest $1$$:$$1$ synchronization is achieved.
As $K$ grows, the oscillations in the two subsystems
become more and more synchronized until at $K$$=$$2$ the two systems
start to fire simultaneously almost all the time.
However, at least in the given range of 
$K$, the perfect $1$$:$$1$ synchronization, with both ISI ratio and 
synchronization index equal to unity, is still not realized. 

\begin{figure}
\epsfxsize=\linewidth
%\epsfbox{FIG/fig_isi_tcor_K_D1_0.6_C_0.2_tau_1.eps}
\epsfbox{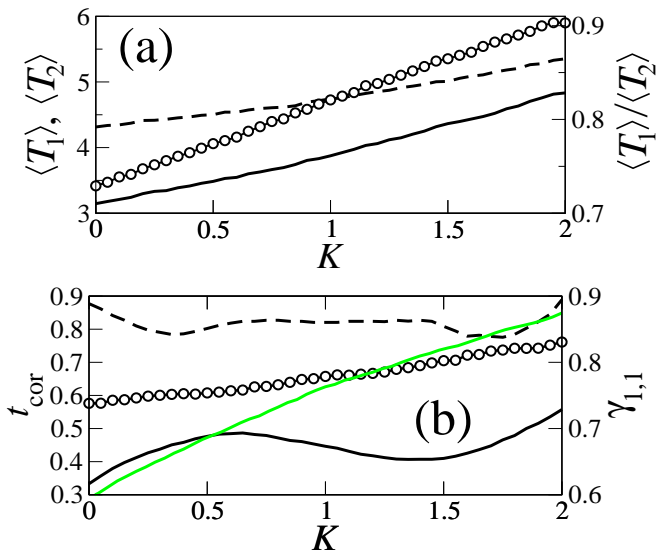}
\caption{\label{fig:ISI_K_D1_0.6_C_0.2_tau_1} (color online)
Timescales, coherence and synchronization index 
of noise-induced oscillations in two coupled
FitzHugh-Nagumo systems Eqs.~(\ref{eqn:fhnc1_fdbk}) and (\ref{eqn:fhnc2})
{\em vs} $K$ of the local feedback at $\tau=1$, 
$D_{1}$$=$$0.6$ and $C$$=$$0.2$ 
(see Fig.~\ref{fig:fhnc_K_tau_D1_0.6_C_0.2} for reference). 
(a) Average ISIs $\langle T_1 \rangle$ (solid line) 
and $\langle T_2 \rangle$ (dashed line), and
their ratio (circles). (b) 
Correlation times $t_{cor}$ obtained from $x_1$ (solid line),
from $x_2$ (dashed line), and from $x_{\Sigma}$ (circles). 
Synchronization index $\gamma_{1,1}$ 
(grey line, green online). 
}\end{figure}

Fig.~\ref{fig:ISI_K_D1_0.6_C_0.2_tau_1}(a) shows the 
full dependences upon $K$ of $\langle T_1 \rangle$ (solid line),
$\langle T_2 \rangle$ (dashed line) and of their ratio (circles). 
In Fig.~\ref{fig:ISI_K_D1_0.6_C_0.2_tau_1}(b) the respective
dependences of correlation time $t_{cor}$ 
from $x_1$ (solid line), $x_2$ (dashed line) 
and from the sum signal $x_{\Sigma}$ (circles), are given together
with the synchronization index $\gamma_{1,1}$ (grey line, green online). 
Both $\langle T_1 \rangle$ and 
$\langle T_2 \rangle$ grow monotonically with $K$, which means
that at $\tau$$=$$1$ the feedback slows down the oscillations.
Within the accuracy of numerical simulation, 
both ISI ratio and $\gamma_{1,1}$
grow linearly with $K$, still not
achieving the value of $1$ at $K$$=$$2$. 
$t_{cor}$ from $x_{\Sigma}$ grows with 
$K$ almost linearly as well, which means that the global
dynamics of the system becomes more ordered with the
stronger feedback.
However, $t_{cor}$ computed from $x_1$ and $x_2$
separately are nonmonotonic.

\begin{figure}
\epsfig{file=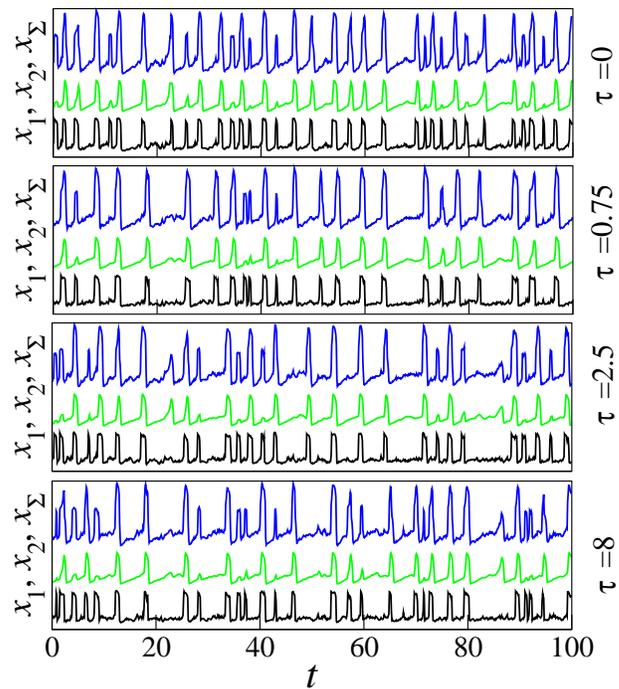,angle=0,width=8cm}
\caption{\label{fig:fig_K_1.5}(color online) 
Realizations of $x_{1}$ (lower), $x_{2}$ (middle) and $x_{\Sigma}$ (upper trace)
of noise-induced oscillations in two coupled
FitzHugh-Nagumo systems Eqs.~(\ref{eqn:fhnc1_fdbk}) and (\ref{eqn:fhnc2})
at $D_1$$=$$0.6$ and $C$$=$$0.2$, 
subject to delayed feedback with $K$$=$$1.5$
for different values of the time delay $\tau$.
See Fig.~\ref{fig:ISI_tau_D1_0.6_C_0.2_K_1.5} 
for reference.}\end{figure}

\begin{figure}
\epsfxsize=\linewidth
%\epsfbox{FIG/fig_isi_tcor_tau_D1_0.6_C_0.2_K_1.5.eps}
\epsfbox{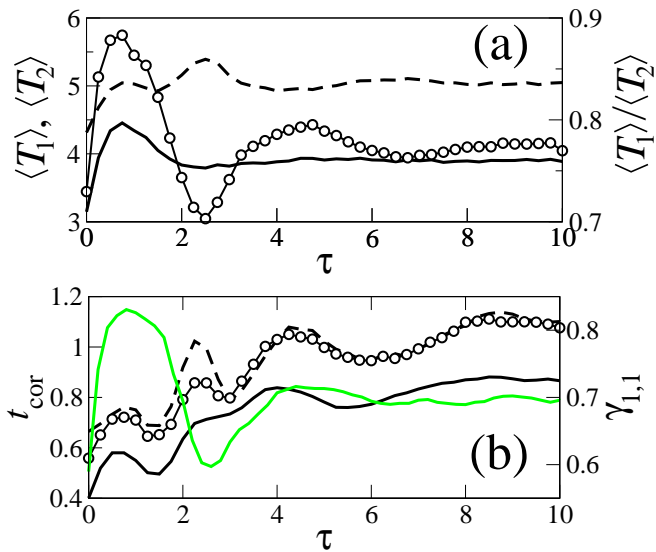}
\caption{\label{fig:ISI_tau_D1_0.6_C_0.2_K_1.5}
Timescales, coherence and synchronization index 
of noise-induced oscillations in two coupled
FitzHugh-Nagumo systems Eqs. (\ref{eqn:fhnc1_fdbk}) and (\ref{eqn:fhnc2})
{\em vs} time delay $\tau$ at $K=1.5$, 
$D_{1}$$=$$0.6$ and $C$$=$$0.2$ 
(see Fig.~\ref{fig:fhnc_K_tau_D1_0.6_C_0.2} for reference). 
Symbols as in Fig. \ref{fig:ISI_K_D1_0.6_C_0.2_tau_1}.
}\end{figure}

Next, we follow the route with a constant $K$$=$$1.5$ 
that crosses the area with the strongest synchronization, 
by changing $\tau$. 
Four respective 
realizations from the subsystems are shown in Fig.~\ref{fig:fig_K_1.5}.
A full picture showing ISIs, their ratio, correlation times and 
synchronization index 
{\it vs} $\tau$ is given in Fig.~\ref{fig:ISI_tau_D1_0.6_C_0.2_K_1.5}.
Both $\langle T_1 \rangle$ and $\langle T_2 \rangle$, as well as their
ratio, change nonmonotonically with
$\tau$ while its value is smaller than $8$. At
$\tau > 8$ they start to 
asymptotically tend to some constant values that are slightly larger 
than those without feedback. 

An increase of $\tau$ from zero leads to an increase of
both $\langle T_1 \rangle$ and $\langle T_2 \rangle$. But 
 $\langle T_1 \rangle$  grows faster than $\langle T_2 \rangle$,
 thus their ratio $\langle T_1 \rangle$$/$$\langle T_2 \rangle$
 grows with $\tau$, as well as the phase synchronization index $\gamma_{1,1}$.
 At the same time, the coherence of each subsystem and
 of their global dynamics grows, too, as illustrated by the
 behavior of the respective correlation times $t_{cor}$
 (Fig.~\ref{fig:ISI_tau_D1_0.6_C_0.2_K_1.5}(b)).
 
 After the maximum of $\langle T_1 \rangle$ and 
 $\langle T_2 \rangle$, and of their ratio, 
 is achieved at $\tau$$\approx$$0.7$,
 both $\langle T_1 \rangle$ and $\langle T_2 \rangle$ start to
 decrease, but again $\langle T_1 \rangle$ decreases faster than
 $\langle T_2 \rangle$, thus their ratio decreases.
 A similar behavior is observed in $t_{cor}$ and in $\gamma_{1,1}$.
 
 Starting from $\tau$$\approx$$2$, the ISI $\langle T_1 \rangle$ 
 of the first system hardly changes with $\tau$. However, counterintuitively,
 the ISI $\langle T_2 \rangle$ of the second system 
 responds to the further increase of $\tau$
 by displaying a noticeable  maximum at $\tau$$\approx$$2.5$.
 This leads to a well-pronounced minimum of the ISI ratio 
 (Fig.~\ref{fig:ISI_tau_D1_0.6_C_0.2_K_1.5}(a)) and
 of the synchronization index $\gamma_{1,1}$ 
(Fig.~\ref{fig:ISI_tau_D1_0.6_C_0.2_K_1.5}(b)). 
 This phenomenon is accompanied by a respective maximum
 of the coherence of the second susbystem and of the global 
 dynamics, while neither 
 the timescales nor the coherence of the first subsystem  change
 substantially. This is a highly counterintuitive phenomenon,
 since the feedback is applied to the first subsystem only, while
 the second subsystem responds to the changes of the feedback
 only indirectly through its coupling with the first subsystem. 

 With the further increase of $\tau$, the dynamics of the second subsystem
 changes more substantially than the one of the first subsystem,
 and thus gives a larger contribution to the changes of
 the global dynamics. 

\subsection{\label{subsec:all_syn}
Control of a weakly, and of a strongly, 
synchronized system}

In this subsection we consider 
subsystems Eqs.~(\ref{eqn:fhnc1_fdbk}) and~(\ref{eqn:fhnc2})
that are either further from ($D_1$$=$$0.6$, $C$$=$$0.1$),
or closer to ($D_1$$=$$0.15$, $C$$=$$0.2$), the
$1$$:$$1$ synchronization region, under the influence
of the controlling feedback.

\begin{figure} 
\epsfxsize=\linewidth
%\epsfbox{FIG/fig_K_tau_D1_0.6_C_0.1.eps}
\epsfbox{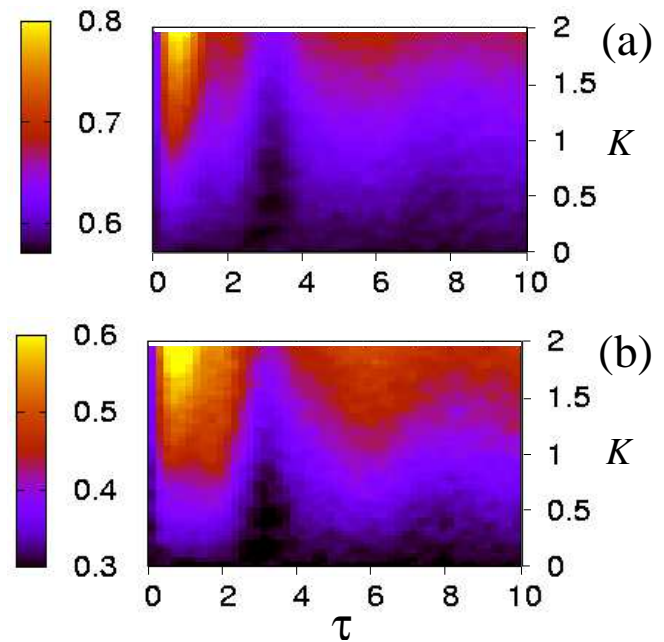}
\caption{\label{fig:fhnc_K_tau_D1_0.6_C_0.1} 
(color online) 
Effect of delayed feedback on frequency and phase 
synchronization betwen the two subsystems at
$D_1$$=$$0.6$ and $C$$=$$0.1$, which are further away
from the $1$$:$$1$ synchronization tongue
shown in Fig.~\ref{fig:fhnc_isi_D1_C},
than those considered in 
Fig.~\ref{fig:fhnc_K_tau_D1_0.6_C_0.2}. 
Plot as in Fig.~\ref{fig:fhnc_K_tau_D1_0.6_C_0.2}. 
}\end{figure}

For $D_1$$=$$0.6$, $C$$=$$0.1$, the 
ratio of ISIs and the synchronization index $\gamma_{1,1}$
are shown by color code 
in Figs.~\ref{fig:fhnc_K_tau_D1_0.6_C_0.1}. 
As with the stronger synchronized subsystems, 
the feedback is able to move the whole system  towards
a more synchronous state. 

\begin{figure}
\epsfxsize=\linewidth
%\epsfbox{FIG/fig_isi_tcor_tau_D1_0.6_C_0.1_K_1.5.eps}
\epsfbox{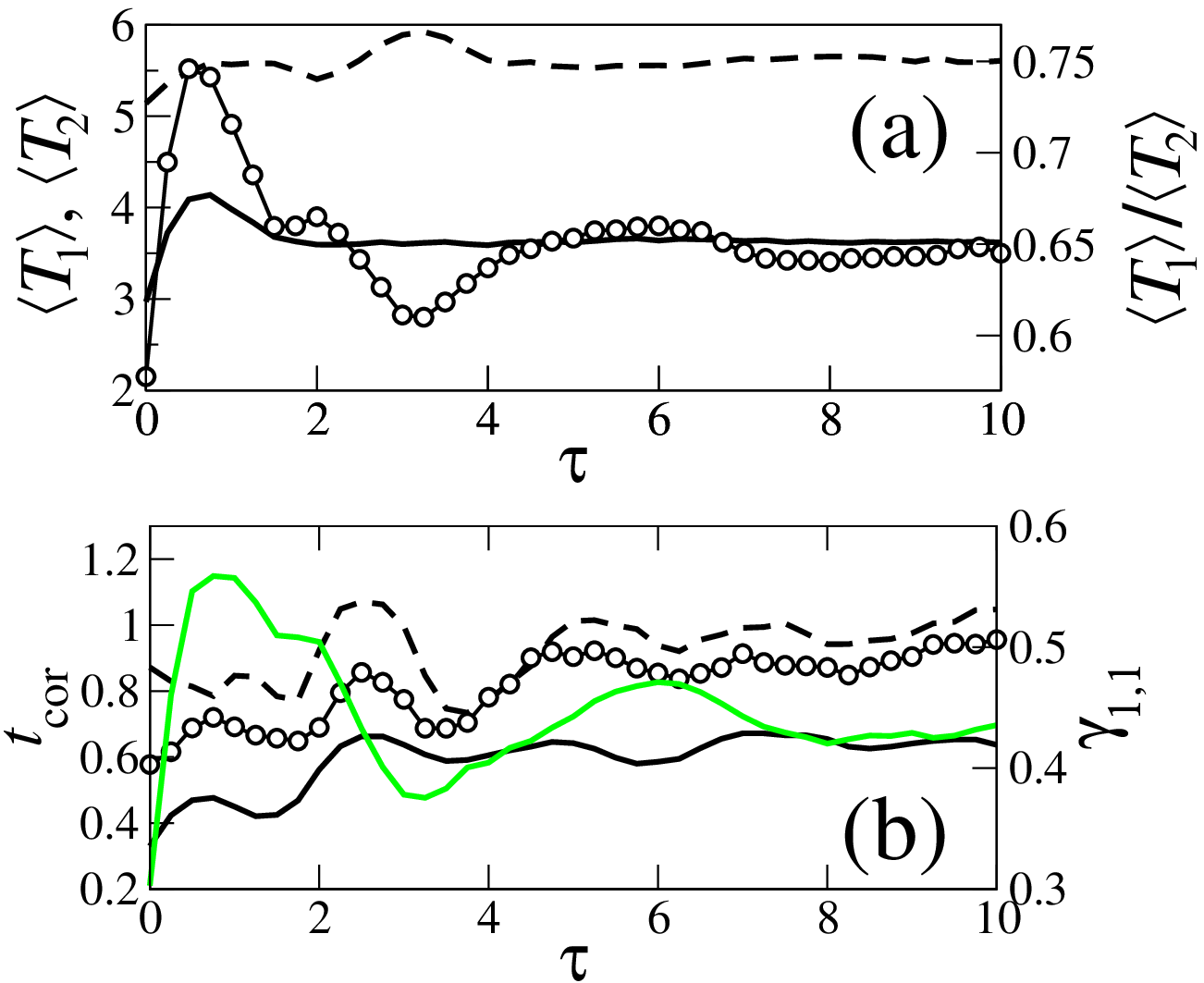}
\caption{\label{fig:ISI_tau_D1_0.6_C_0.1_K_1.5}
Timescales, coherence and synchronization index 
of noise-induced oscillations in two coupled
FitzHugh-Nagumo systems Eqs. (\ref{eqn:fhnc1_fdbk}) and (\ref{eqn:fhnc2})
depending on the time delay $\tau$ at $K=1.5$, 
$D_{1}$$=$$0.6$ and $C$$=$$0.1$ 
(see Fig. \ref{fig:fhnc_K_tau_D1_0.6_C_0.1} for reference). 
Symbols as in Fig. \ref{fig:ISI_K_D1_0.6_C_0.2_tau_1}.
}\end{figure}

As with the example of Sec.~\ref{subsec:moderate_syn},
we consider the cut of the $\tau$--$K$ plane along
$K=1.5$, choosing the route that goes through the lighter area
of the largest synchronization index. 
ISIs, their ratio, correlation times, 
and synchronization index are shown in 
Fig.~\ref{fig:ISI_tau_D1_0.6_C_0.1_K_1.5} depending on
$\tau$. Their behavior has some similarities 
to that in a moderately synchronized system of 
Sec.~\ref{subsec:moderate_syn}.

Namely, the initial increase of $\tau$ from zero leads
to the growth of both  $\langle T_1 \rangle$ and $\langle T_2 \rangle$,
the former growing faster than the latter. This leads to
the growth of the ISI ratio and of synchronization
index $\gamma_{1,1}$, and also 
of $t_{cor}$ of $x_1$ and of $x_{\Sigma}$. 
All variables achieve the maximum at $\tau$$\approx$$0.6$.
After that, all the variables describing the first system
start to decrease, while $\langle T_2 \rangle$ does not change
until $\tau$$=$$1.5$. Here, the ISI ratio decreases correspondingly, 
like for $C$$=$$0.2$. And again, after $\tau$$=$$2$,
$\langle T_1 \rangle$ hardly changes with $\tau$, while
$\langle T_2 \rangle$ exhibits a noticeable maximum
that leads to the rapid drop of both ISI ratio and synchronization index.

With further increase of $\tau$ beyond $6$, both subsystems respond 
only slightly
and comparably to the changes in $\tau$.

Note that either for a moderately synchronized system, or for
a system that is less synchronized, the feedback is able to
make $1$$:$$1$ synchronization substantially 
stronger for a suitable choice of 
its parameters. However, it cannot destroy the existing 
synchronization or weaken it as much as it can strengthen it. 

\begin{figure} 
\epsfxsize=\linewidth
%\epsfbox{FIG/fig_K_tau_D1_0.15_C_0.2.eps}
\epsfbox{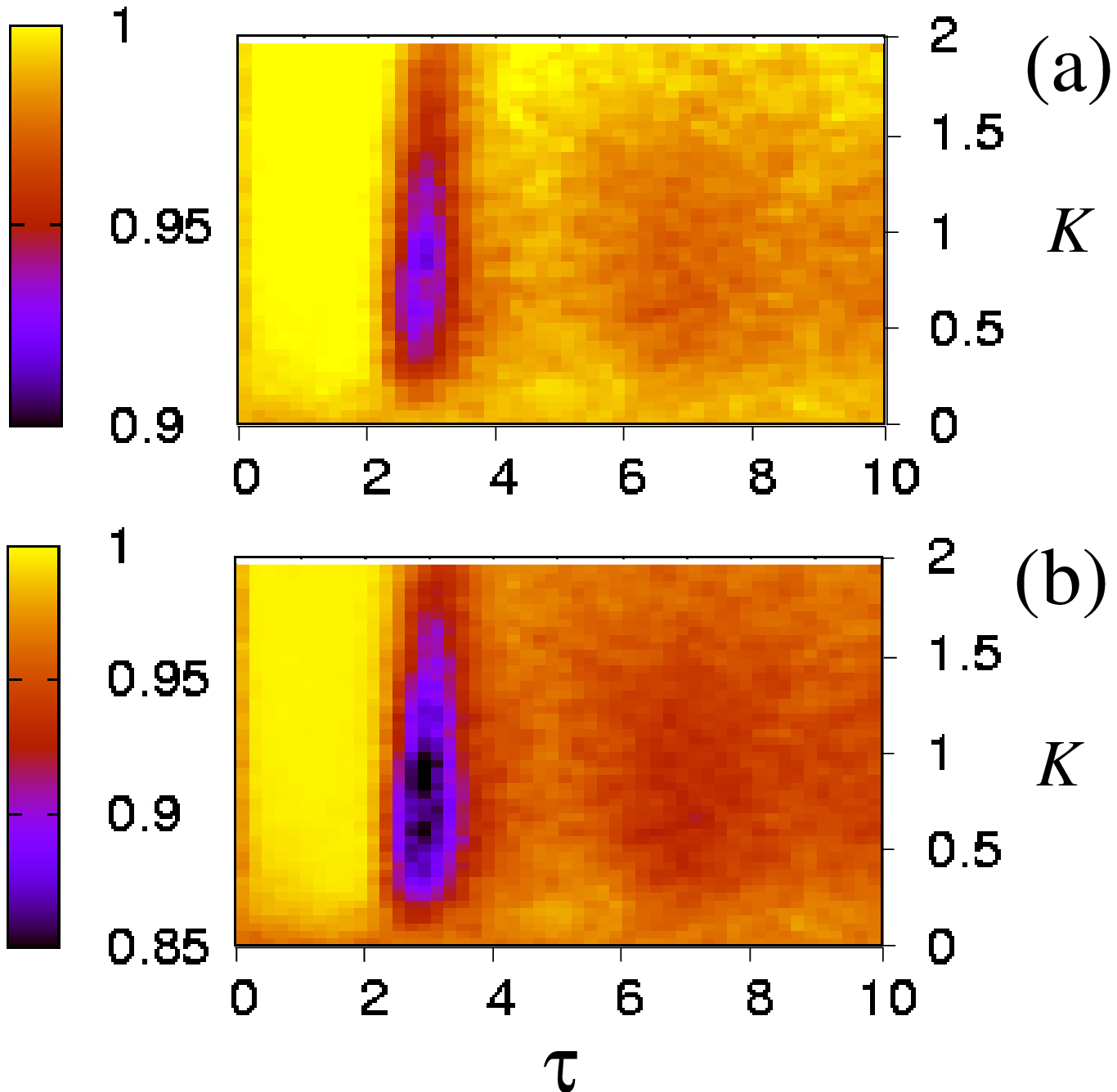}
\caption{\label{fig:fhnc_K_tau_D1_0.15_C_0.2} (color online) 
Effect of delayed feedback on frequency and phase 
synchronization betwen the two subsystems at
$D_1$$=$$0.15$ and $C$$=$$0.2$, which are closer to
the $1$$:$$1$ synchronization tongue 
shown in Fig.~\ref{fig:fhnc_isi_D1_C}, than those considered in
Fig.~\ref{fig:fhnc_K_tau_D1_0.6_C_0.2}. 
Plot as in Fig.~\ref{fig:fhnc_K_tau_D1_0.6_C_0.2}. 
}\end{figure}

For the system that is very well synchronized from the beginning
at $D_1$$=$$0.15$ and $C$$=$$0.2$ with $\gamma_{1,1}$$=$$0.99$, 
we reveal the ISI ratio and synchronization index $\gamma_{1,1}$ 
for a large range of $K$ and $\tau$ 
(Fig.~\ref{fig:fhnc_K_tau_D1_0.15_C_0.2}). Already this picture
shows that delayed feedback can either enhance or suppress
synchronization. 

\begin{figure}
\epsfxsize=\linewidth
%\epsfbox{FIG/fig_isi_tcor_tau_D1_0.15_C_0.2_K_1.eps}
\epsfbox{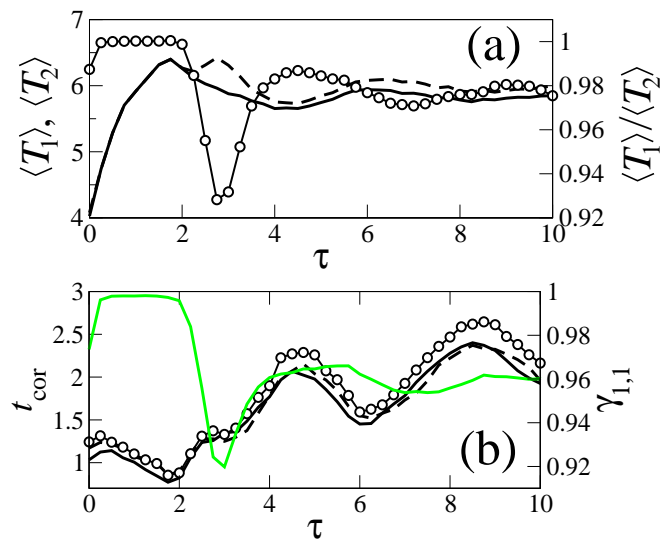}
\caption{\label{fig:ISI_tau_D1_0.15_C_0.2_K_1}
Timescales, coherence and synchronization index 
of noise-induced oscillations in two coupled
FitzHugh-Nagumo systems Eqs. (\ref{eqn:fhnc1_fdbk}) and (\ref{eqn:fhnc2})
depending on the time delay $\tau$ at $K=1$,
$D_{1}$$=$$0.15$ and $C$$=$$0.2$ 
(see Fig. \ref{fig:fhnc_K_tau_D1_0.15_C_0.2} for reference). 
Symbols as in Fig. \ref{fig:ISI_K_D1_0.6_C_0.2_tau_1}.
}\end{figure}

For a more detailed picture of the phenomena induced by the
feedback, a cut of this picture along $K$$=$$1$ is given 
in Fig.~\ref{fig:ISI_tau_D1_0.15_C_0.2_K_1} where the ISIs
and their ratio are shown, together with $\gamma_{1,1}$ 
and correlation times for $x_1$, $x_2$ and $x_{\Sigma}$. 
An immediate obvious observation is that, in contrast to the
two previously considered cases of less synchronous subsystems,
here the feedback can make synchronization perfect with
$\gamma_{1,1}$$=$$1$, and can maintain it like this for
a substantial range of $\tau$$\in$$[0.25; 2]$
(Fig.~\ref{fig:ISI_tau_D1_0.15_C_0.2_K_1}(a)). 
The fact that the two subsystems are very synchronous 
from the beginning is
also supported by very similar values of the correlation times of
both systems' realizations and of their sum at $\tau$$=$$0$
(Fig.~\ref{fig:ISI_tau_D1_0.15_C_0.2_K_1}(b)).

As $\tau$ is slighly increased from zero, as in the two
previous examples, 
both $\langle T_1 \rangle$ and $\langle T_2 \rangle$
grow. But, as before, $\langle T_1 \rangle$ grows 
a little faster than $\langle T_2 \rangle$.
This can hardly be resolved in the plots, since 
$\langle T_1 \rangle$  and $\langle T_2 \rangle$
are very close and hardly distinguishable. However,
the difference between them, and the disapperanace of
this difference, is visible through the ISI ratio
(Fig.~\ref{fig:ISI_tau_D1_0.15_C_0.2_K_1}(a)).

With this, $t_{cor}$ and $\gamma_{1,1}$ slightly 
grow, too. All quantities considered achieve their maxima
at $\tau$$\approx$$0.25$. After that, as $\tau$ increases,
the individual ISIs $\langle T_1 \rangle$  and $\langle T_2 \rangle$
grow simultaneously and remain equal, so that their ratio
and synchronization index $\gamma_{1,1}$ stay equal to $1$
with high accuracy 
in the range $\tau$$\in$$[0.25; 2]$. 

However, surprisingly, 
while the ISI ratio and $\gamma_{1,1}$ are equal to $1$, 
i.e., the subsystems
maintain the same level of perfect synchrony, 
all three correlation times decrease with $\tau$. 
This means that while the two subsystems fire simultaneously,
these firings occur less regularly. Thus, the feedback here can 
introduce disorder into the system without destroying
its perfect synchronization. 

Then, as $\tau$ continues to increase beyond the value of $2$,
as in the previous examples, the second subsystem demonstrates
a noticeable maximum of its ISI $\langle T_2 \rangle$. 
Although unlike in the two other examples, here 
$\langle T_1 \rangle$
continues to decrease, the ISI ratio and $\gamma_{1,1}$ 
exhibit a sharp minimum, and then grow again. 

At $\tau$ increases beyond $8$, the ISI ratio and $\gamma_{1,1}$ 
become less and less dependent on $\tau$, asymptotically
tending to some values that are only slighly lower than without
the feedback. On the contrary, $t_{cor}$ continues to
change nonmonotonically with $\tau$, never becoming less than without
the feedback. Since the system is in the state of a strong 
synchronization throughout the changes in $\tau$, the changes
in all three curves $t_{cor}$ occur synchronously. 

Thus, the feedback can make both local and global
dynamics of the system more coherent, and at the same time 
weaken synchronization. 

\section{\label{sec:sum}Summary 
and Conclusions}

We have considered a model describing two coupled excitable neurons, 
in the form of two mutually coupled non-identical excitable
FitzHugh-Nagumo systems, subject to independent sources of noise 
with different strengths. 
In order to assess the effect of time-delayed feedback control upon 
the coupled system, we have analyzed the following characteristics:
the timescales of the individual systems quantified as mean interspike 
intervals (ISI); the ISI ratio as a measure of frequency
synchronization; the coherence quantified by the
correlation time of the individual subsystems' realizations and of
their sum; and the index of $1$$:$$1$ phase synchronization
between the subsystems. 

The coupled system without control displays a $1$$:$$1$ 
synchronization tongue in the $(D_1, C)$ parameter plane, given by  
the noise strength $D_1$ in the first subsystem and the
coupling strength $C$. Interestingly, frequency 
and phase synchronization occurred in the same area of the parameter plane. 
Two mechanisms for synchronization were identified:
phase (frequency) locking, and suppression of natural dynamics, respectively. 

Next, the first of the two interacting subsystems was subjected 
to the local delayed feedback with the aim to manipulate the
global dynamics of the system of interacting oscillators.
The feedback force was constructed as a difference between the
current state of the system and its state some $\tau$ time units before,
multiplied by a positive constant $K$. 

The delayed feedback was applied to the system in three states
of synchrony: moderately synchronized, weakly synchronized,
and strongly synchronized. In all three cases, $1$$:$$1$ synchronisation could
be either improved or weakened, depending upon the choice of 
$\tau$ and $K$. Like the correlation times, the synchronisation index is
modulated nonmonotonically as a function of the delay time $\tau$,
indicating that there is resonance-like behavior for certain values of $\tau$.
Perfect synchronisation can only be achieved if the uncontrolled state is already
sufficiently synchronized. 

The mechanism behind the reported action of the delayed feedback 
is as follows. As it was shown earlier \cite{JAN04,BAL04},
the feedback applied to a single excitable system is able
to change the timescales and coherence of noise-induced 
oscillations. When the system subjected to the feedback 
is coupled to another system, the shift of the timescale
of the former will lead to a proportional shift of the timescale
of the latter. The exact magnitude of the shift in the second subsystem
will depend on the closeness of the two subsystems to the
state of synchronization. Only if the two subsystems
are sufficiently $1$$:$$1$ synchronized from the beginning,
the shift in the second system can be expected to match the
shift in the first system. 

Interestingly, the above mechanism does not always work
in the system considered. Namely, for some ranges of time delay
$\tau$, the change in $\tau$ does not cause any noticeable
change in the system to which the feedback is applied. However,
it does change the properties of oscillations in the system
that is coupled to it, albeit that does not experience the influence
of the feedback directly. 

An important observation is that the delay-induced 
increase of coherence of the global dynamics 
is most frequently accompanied by the growth of 
the degree of synchronization.
However, 
a high synchonization index does 
not always mean high coherence: 
delayed feedback can induce, or make stronger,
the synchronization between the two subsystems,
but the state of each subsystem, and their global
dynamics, can  become more disordered at the same time.
The converse is also true.

It is remarkable that delayed feedback control can influence global characteristics 
of the two coupled neurons although the control is only applied locally to a subsystem.
We were able to enhance or destroy the regularity of oscillations and the 
stochastic synchronization of the two neurons by choosing appropriate control parameters, 
in particular a suitable delay time.

We consider these findings as important for the understanding of coupled 
nonlinear systems and see possible applications especially in neuroscience. 
In fact, experimental studies of two coupled neurons from the 
stomatogastric ganglion of a lobster \cite{PIN00}, and from a leech 
\cite{DEM01} have reported various degrees of synchrony of 
excitatory postsynaptic potentials.  As stochastic sources of the 
spontaneous random firing of neurons, noise due to the 
conducting ion channels, synaptic noise, and noise resulting from the 
coupling to a large number of other neurons emitting signals, have been 
identified \cite{ZEN04}. 
Also it was demonstrated experimentally \cite{JUN98} that   
spatially and temporally coherent $Ca^{2+}$ waves, mediated by network 
noise, may play an important role in generating correlated neural activity.
By applying delayed feedback control to real neural systems one should 
be able to influence neural synchrony. 
First results of applying time-delayed neurofeedback from realtime MEG 
signals to humans via visual stimulation in order to 
suppress the alpha rhythm, which is observed due to strongly synchronized
neural populations in the visual cortex in the brain, look promising 
\cite{TASS06}.
Further work should focus on more 
sophisticated models and on coupling more than two neurons.

\section{Acknowledgments}
This work was partially supported by DFG in the framework of Sfb 555
(Complex nonlinear processes). 
B.H. thanks the Studienstiftung des deutschen Volkes for support and
 gratefully 
acknowledges the hospitality of Loughborough 
University. A.B. acknowledges the support of EPSRC (UK).

%\bibliographystyle{unsrt} %prsty-fullauthor
%\bibliography{ref}

\begin{thebibliography}{10}

\bibitem{BRA94}
H.~A. Braun, H.~Wissing, K.~Sch{\"a}fer, and M.~C. Hirsch, 
%\newblock Oscillation and noise determine signal transduction in shark
%  multimodal sensory cells.
\newblock {\em Nature} {\bf 367}, 270 (1994).

\bibitem{LU95}
J.~Lu and H.~M. Fishman,
%\newblock Localization and function of the electrical oscillation in
%  electroreceptive ampullary epithelium from skates.
\newblock {\em Biophys.~J.} {\bf 69}, 2458 (1995).

\bibitem{EGU00}
V.~M. Eguiluz, M.~Ospeck, Y.~Choe, A.~J. Hudspeth, and M.~O. Magnasco,
%\newblock Essential nonlinearities in hearing.
\newblock {\em Phys.~Rev.~Lett.} {\bf 84}, 5232 (2000).

\bibitem{SOF93}W.R.~Softky, C. Koch, {\em J. Neurosci.} {\bf 13}, 334 (1993).

\bibitem{HOD52} A.L.~Hodgkin and A.F.~Huxley, 
{\em J. Physiol. London} {\bf 117}, 500 (1952).

\bibitem{JUN98}P.~Jung, A.~Cornell-Bell, K.S.~Madden, 
and F.~Moss, {\em J. Neurophysiol.}
 {\bf 79}, 1098 (1998). 

\bibitem{BAD05} M.~Badoual, M.~Rudolph, Z.~Piwkowska, A.~Destexhe, 
and T.~Bal, 
{\em Neurocomp.} {\bf 65}, 493 (2005).

\bibitem{WAN94}
X.-J. Wang,
\newblock {\em Neuroscience} {\bf 59}, 21 (1994).

\bibitem{NEI01}
A.~Neiman and D. F. Russell,
%\newblock Stochastic biperiodic oscillations in the electroreceptors of
%  paddlefish.
\newblock {\em Phys.~Rev.~Lett.} {\bf 86}, 3443 (2001).

\bibitem{SAM04} J.M.~Samonds, J.D.~Allison, H.A.~Brown, 
and A.B. Bonds, {\em Proc. Natl. Acad. Sci. USA} {\bf 101}, 6722 (2004).

\bibitem{BEN04} A.~Benucci, P. F. M. J.~Verschure, and P.~K{\"o}nig, 
{\em Phys.~Rev. E }
{\bf 70}, 051909 (2004). 

\bibitem{TAS98}
P.~Tass, M.~G. Rosenblum, J.~Weule, J.~Kurths, A.~Pikovsky, J.~Volkmann,
  A.~Schnitzler, and H.-J. Freund,
%\newblock Detection of n:m phase locking from noisy data: Application to
%  magnetoencephalography.
\newblock {\em Phys.~Rev.~Lett.} {\bf 81}, 3291 (1998).

\bibitem{GRO02} P. Grosse, M. J. Cassidy, P. Brown, 
{\em Clinical Neurophysiology} {\bf 113}, 1523 (2002).

\bibitem{OTT90}
E.~Ott, C.~Grebogi, and J.~A. Yorke,
%\newblock Controlling chaos.
\newblock {\em Phys.~Rev.~Lett.} {\bf 64}, 1196 (1990).

\bibitem{SCH99c} H. G. Schuster (Ed.),
{\it Handbook of Chaos Control}  (Wiley-VCH, Weinheim, 1999).

\bibitem{BOC00} S. Boccaletti, C.Grebogi, Y. C. Lai, H. Mancini, and D. Maza,
%"The control of chaos: theory and applications",
{\em Physics Reports} {\bf 328}, 103 (2000).

\bibitem{ROS04}
M.~G. Rosenblum and A.~S. Pikovsky,
%\newblock Delayed feedback control of collective synchrony: An approach to
%  suppression of pathological brain rhythms.
\newblock {\em Phys.~Rev.~E} {\bf 70}, 041904 (2004).

\bibitem{POP05} O.V.~Popovych, C.~Hauptmann, P.A.~Tass, 
%Effective desynchronization by nonlinear delayed feedback 
{\em Phys. Rev. Lett.} {\bf 94}, 164102 (2005).

\bibitem{PYR92}
K.~Pyragas,
%\newblock Continuous control of chaos by self-controlling feedback.
\newblock {\em Phys.~Lett.~A} {\bf 170}, 421 (1992).

\bibitem{BAB02} N. Baba, A. Amann, E. Sch{\"o}ll, and W. Just,
%"Giant improvement of time-delayed feedback control by
%        spatio-temporal filtering", 
{\it Phys. Rev. Lett.}
{\bf 89}, 074101 (2002).

\bibitem{BEC02} O. Beck, A. Amann, E. Sch{\"o}ll, J.E.S. Socolar, 
and W. Just,
%"Comparison of time-delayed feedback schemes for
%spatio-temporal control of chaos in a reaction-diffusion system with
%global coupling", 
{\it Phys. Rev. E} {\bf 66}, 016213 (2002).

\bibitem{UNK03} 
J. Unkelbach, A. Amann, W. Just, and E. Sch{\"o}ll,
%"Time--delay autosynchronization of the
%spatio-temporal dynamics in resonant tunneling
%diodes",  
{\it Phys. Rev. E} {\bf 68}, 026204 (2003).

\bibitem{JAN04}
N.~B. Janson,  A.~G. Balanov, and E.~Sch{\"o}ll,
%\newblock 	Delayed Feedback as a Means of Control of Noise-Induced Motion.
\newblock {\em Phys. Rev. Lett.} {\bf 93}, 010601 (2004).

\bibitem{BAL04}
A.~G. Balanov, N.~B. Janson, and E.~Sch{\"o}ll,
%\newblock Control of noise-induced oscillations by delayed feedback.
\newblock {\em Physica~D} {\bf 199}, 1 (2004).

\bibitem{POM05a}
J.~Pomplun, A.~Amann, and E.~Sch{\"o}ll,
%\newblock Mean field approximation of time-delayed feedback control of
%  noise-induced oscillations in the {V}an der {P}ol system.
\newblock {\em Europhys.~Lett.} {\bf 71}, 366 (2005).

\bibitem{SAN06} 
G. J. Escalera Santos, J. Escalona, and P. Parmananda, 
{\it Phys. Rev. E} {\bf 73}, 042102 (2006).

\bibitem{LIN04}
B.~Lindner, J.~Garc{\'{\i}}a-Ojalvo, A.~Neiman, and L.Schimansky-Geier,
%\newblock Effects of noise in excitable systems.
\newblock {\em Phys.~Rep.} {\bf 392}, 321(2004).

\bibitem{LIL94} 
D. T. J. Liley and J. J. Wright,
{\it Network Comp. Neur. Systems} {\bf 5}, 175 (1994).

\bibitem{PIN00} 
R. D. Pinto, P. Varona, A. R. Volkovskii, A. Sz{\"u}cs, H. D. I. Abarbanel 
and M. I. Rabinovich,
{\it Phys. Rev. E} {\bf 62}, 2644 (2000).

\bibitem{DEM01} 
F. F. De-Miguel, M. Vargas-Caballero, and E. Garcia-Perez,
{\it J. Exp. Biol.} {\bf 204}, 3241 (2001).

\bibitem{Ros_Moss_book} M.G. Rosenblum, A.S. Pikovsky, J. Kurths, 
C. Sch\"{a}fer, and P. Tass,
%"Phase Synchronization: From Theory to Data Analysis,"
In: {\em Handbook of Biological Physics} (Elsevier Science, Amsterdam, 2001), 
%Series Editor  A.J. Hoff,
Vol. 4,  {\em Neuro-informatics and Neural Modeling},  
Eds. F. Moss and S. Gielen,
Ch. 9,  279-321.

\bibitem{Lind_CR_99} B. Lindner, L. Schimansky-Geier, 
%Analytical approach to the stochastic
%FitzHugh-Nagumo system and coherence resonance, 
{\em Phys. Rev. E} {\bf 60}, 7270 (1999).

\bibitem{PIK97}
A.S. Pikovsky and J.~Kurths,
%\newblock Coherence resonance in a noise-driven excitable system.
\newblock {\em Phys.~Rev.~Lett.} {\bf 78}, 775 (1997).

\bibitem{HAN99}S.K.~Han, T.G.~Yim, D.E.~Postnov, 
and O.V.~Sosnovtseva, {\em Phys.~Rev.~Lett.} {\bf 83}, 1771 (1999).

\bibitem{POS02}
D.~E. Postnov, O.~V. Sosnovtseva, S.~K. Han, and W.~S. Kim,
%\newblock Noise-induced multimode behavior in excitable systems.
\newblock {\em Phys.~Rev.~E} {\bf 66}, 016203 (2002).

\bibitem{LAN80} P. Landa, 
{\it Self-oscillations in the systems with finite 
number of degree of freedom} (Nauka, Moscow, 1980) (in Russian).

\bibitem{MOS02} 
E. Mosekilde, Yu. Maistrenko, and D. Postnov, 
{\it Chaotic synchronization. Applications to Living Systems}, 
World Scientific, Singapore, Series A, Vol. 42 (2002).

\bibitem{ZEN04}
S. Zeng and P. Jung, {\em Phys. Rev. E} {\bf 70}, 011903 (2004). 

\bibitem{TASS06}
V. Hadamschek, {\em PhD Thesis}, TU Berlin (2006).



\end{thebibliography}
%%%%%%%%%%%%%%%%%%%%%%%%%%%%%%%%%%%%%%%%%%%%%%%%%%%%%%%%%%%%%%%%%%%%%

\end{document}